\documentclass[a4paper]{jpconf}

\usepackage{graphicx,color}
\usepackage{amsmath,amssymb,bm}

\newcommand{\ls}{\left[}
\newcommand{\rs}{\right]}

\newcommand{\ff}[1]{\frac{1}{#1}}

\begin{document}

\title{Pseudospin symmetry: Recent progress with supersymmetric quantum mechanics}

\author{Haozhao Liang$^{1,2,*}$, Jie Meng$^{2,3}$, Shihang Shen$^2$, Nguyen Van Giai$^4$, Shuangquan Zhang$^{2}$, Ying Zhang$^{2,5}$, and Pengwei Zhao$^2$}

\address{$^{1}$RIKEN Nishina Center, Wako 351-0198, Japan}
\address{$^{2}$School of Physics, Peking University, Beijing 100871, China}
\address{$^{3}$School of Physics and Nuclear Energy Engineering, Beihang University, Beijing 100191, China}
\address{$^{4}$Institut de Physique Nucl\'eaire, IN2P3-CNRS, Universit\'e Paris-Sud, F-91406 Orsay, France}
\address{$^{5}$Graduate School of Science and Technology, Niigata University, Niigata 950-2181, Japan}

\ead{haozhao.liang@riken.jp}

\begin{abstract}
  It is an interesting and open problem to trace the origin of the pseudospin symmetry in nuclear single-particle spectra and its symmetry breaking mechanism in actual nuclei.
  In this report, we mainly focus on our recent progress on this topic by combining the similarity renormalization group technique, supersymmetric quantum mechanics, and perturbation theory.
  We found that it is a promising direction to understand the pseudospin symmetry in a quantitative way.
\end{abstract}

\section{Introduction}

Although the concept of pseudospin symmetry (PSS) in nuclear single-particle spectra has been introduced for more than 40 years, the origin of PSS and its breaking mechanism in realistic nuclei have not been fully understood. Basically, whether or not its nature is perturbative remains an open problem.
In this report, we mainly focus on our recent progress on this topic by combining the similarity renormalization group (SRG) technique, supersymmetric (SUSY) quantum mechanics, and perturbation theory \cite{Liang2013,Shen2013}.

The PSS was introduced in 1969 to explain the near degeneracy between pairs of single-particle states with the quantum numbers ($n-1, l + 2, j = l + 3/2$) and ($n, l, j=l + 1/2$) \cite{Arima1969,Hecht1969}.
They are regarded as the pseudospin doublets with modified quantum numbers ($\tilde{n}=n-1,\tilde{l}=l+1,j=\tilde{l}\pm1/2$).
Since the suggestion of PSS, there have been comprehensive efforts to understand its origin.
In 1997, the PSS was shown to be a relativistic symmetry of the Dirac Hamiltonian, and the equality in magnitude but difference in sign of the scalar $\mathcal{S}(\bm{r})$ and vector $\mathcal{V}(\bm{r})$ potentials was suggested as the exact PSS limit \cite{Ginocchio1997}.
A more general condition $d(\mathcal{S}+\mathcal{V})/dr=0$ \cite{Meng1998,Sugawara-Tanabe1998} can be better satisfied in exotic nuclei with diffuse potentials \cite{Meng1999}.
On the other hand, since there exist no bound nuclei within such PSS limit, the non-perturbative or dynamical nature of PSS in realistic nuclei was suggested \cite{Alberto2001}.
A recent overview on the PSS investigation can be found in Ref.~\cite{Liang2013}, and the readers are referred to Refs.~\cite{Lu2012,Jolos2012,Alberto2013} for some recent progress.

Recently, the perturbation theory was used to investigate the symmetries of the Dirac Hamiltonian and their breaking in realistic nuclei~\cite{Liang2011,Li2011}, which provides a clear and quantitative way for investigating the perturbative nature of PSS.
On the other hand, the SUSY quantum mechanics can provide a PSS-breaking potential without singularity~\cite{Typel2008}, and naturally interpret the unique feature that all states with $\tilde{l}>0$ have their own pseudospin partners except for the intruder states~\cite{Typel2008,Leviatan2004}.
Furthermore, the SRG technique fills the gap between the perturbation calculations and the SUSY descriptions by transforming the Dirac Hamiltonian into a diagonal form which keeps every operator Hermitian~\cite{Guo2012,Li2013}.
Therefore, we deem it promising to understand the PSS and its breaking mechanism in a quantitative way by combining the SRG technique, SUSY quantum mechanics, and perturbation theory \cite{Liang2013,Shen2013}.

\section{SRG, SUSY quantum mechanics, and perturbation theory}

Within the relativistic scheme, our starting point is the Dirac Hamiltonian for nucleons, $H_D = \bm{\alpha\cdot p} + \beta (M+\mathcal{S}) + \mathcal{V}$, where $\bm{\alpha}$ and $\beta$ are the Dirac matrices, $M$ is the nucleon mass, $\mathcal{S}$ and $\mathcal{V}$ are the scalar and vector potentials, respectively.

Using the SRG technique \cite{Guo2012}, the Dirac Hamiltonian can be transformed into a series of $1/M$ having a diagonal form.
By keeping the leading-order terms of the kinetic energy, the central and spin-orbit (SO) potentials, respectively, the eigenequations for nucleons in the Fermi sea read
\begin{equation}\label{Eq:Schrodinger}
    \left[-\frac{1}{2M}\frac{d^2}{dr^2} + \frac{\kappa(\kappa+1)}{2Mr^2} + V(r) + \frac{\kappa}{Mr}U(r) \right] R(r) = E R(r)
\end{equation}
with $V(r) = \mathcal{V}+\mathcal{S}$ and $U(r)=-(\mathcal{V}-\mathcal{S})'/(4M)$.
Here the spherical symmetry is adopted, the symbol $'$ means the derivative with respect to $r$, and the good quantum number $\kappa$ is defined as $\kappa=\mp(j+1/2)$ for $j=l\pm1/2$.
One finds that Eq.~(\ref{Eq:Schrodinger}) is nothing but the Schr\"odinger equation including a SO term.

In the SUSY framework \cite{Cooper1995}, a couple of Hermitian conjugate first-order differential operators are defined as
\begin{equation}
B_\kappa^+ = \ls Q_\kappa(r)-\frac{d}{dr}\rs \frac{1}{\sqrt{2M}},\qquad B_\kappa^- = \frac{1}{\sqrt{2M}}\ls Q_\kappa(r)+\frac{d}{dr}\rs,
\end{equation}
where the $Q_\kappa(r)$ are the so-called superpotentials to be determined.
In order to explicitly identify the $\kappa(\kappa+1)$ structure and the SO term shown in Eq.~(\ref{Eq:Schrodinger}), one can introduce the reduced superpotentials $q_\kappa(r) = Q_\kappa(r) - \kappa/r - U(r)$.
In such a way, the SUSY partner Hamiltonians $H_1$ and $H_2$ can be expressed as
\begin{subequations}\label{Eq:BB}
\begin{align}
    H_1(\kappa) &= B^+_\kappa B^-_\kappa = \ff{2M}\ls-\frac{d^2}{dr^2}+\frac{\kappa(\kappa+1)}{r^2}+q_\kappa^2+2q_\kappa U+U^2+\frac{2\kappa}{r}q_\kappa-q'_\kappa-U'\rs+\frac{\kappa}{Mr}U,\notag \\ \label{Eq:BB1}\\
    H_2(\kappa) &= B^-_\kappa B^+_\kappa = \ff{2M}\ls-\frac{d^2}{dr^2}+\frac{\kappa(\kappa-1)}{r^2}+q_\kappa^2+2q_\kappa U+U^2+\frac{2\kappa}{r}q_\kappa+q'_\kappa+U'\rs+\frac{\kappa}{Mr}U.\notag \\ \label{Eq:BB2}
\end{align}
\end{subequations}
Note that the Hamiltonian $H(\kappa)$ in Eq.~(\ref{Eq:Schrodinger}) and its SUSY partner Hamiltonian $\tilde{H}(\kappa)$ differ from $H_1$ and $H_2$ by a common constant $e(\kappa)$, the so-called energy shift \cite{Liang2013,Typel2008,Cooper1995}.
By combining Eqs.~(\ref{Eq:Schrodinger}) and (\ref{Eq:BB1}), one obtains the first-order differential equations for the reduced superpotentials $q_\kappa(r)$, and then obtains $\tilde{H}(\kappa)$ with Eq.~(\ref{Eq:BB2}).
Now it is important to see that not only does the $\kappa(\kappa+1)$ structure appear in $H$ ($H_1$) but also the $\kappa(\kappa-1)$ structure explicitly appears in the SUSY partner Hamiltonian $\tilde{H}$ ($H_2$).
The so-called pseudo-centrifugal barrier (PCB) terms $\kappa(\kappa-1)/(2Mr^2)$ are identical for the pseudospin doublets $a$ and $b$ with $\kappa_a + \kappa_b = 1$.

For the perturbation analysis \cite{Liang2011}, the Hamiltonian $\tilde{H}$ is further expressed as $\tilde{H} = \tilde{H}^{\rm PSS}_0 + \tilde{W}^{\rm PSS}$, where $\tilde{H}^{\rm PSS}_0$ and $\tilde{W}^{\rm PSS}$ are the PSS-conserving and PSS-breaking terms, respectively. By requiring that $\tilde{W}^{\rm PSS}$ should be proportional to $\kappa$~\cite{Liang2013,Shen2013}, which is similar to the case of the SO term in the normal scheme, one has
\begin{equation}
    \tilde{H}^{\rm PSS}_0=\ff{2M}\ls-\frac{d^2}{dr^2}+\frac{\kappa(\kappa-1)}{r^2}\rs + \tilde{V}_{\rm PSS}(r),\qquad
    \tilde{W}^{\rm PSS}=\kappa \tilde{V}_{\rm PSO}(r).
\end{equation}
Finally, for a given pair of pseudospin doublets, the pseudospin-orbit (PSO) potential $\tilde{V}_{\rm PSO}(r)$ can be uniquely determined as
\begin{equation}\label{Eq:VPSO}
  \tilde{V}_{\rm PSO}(r)
        =\frac{1}{M}\frac{q'_{\kappa_a}(r) - q'_{\kappa_b}(r)}{\kappa_a-\kappa_b}
        + \frac{1}{Mr}U(r).
\end{equation}

\begin{figure}\centering
\includegraphics[width=0.45\textwidth]{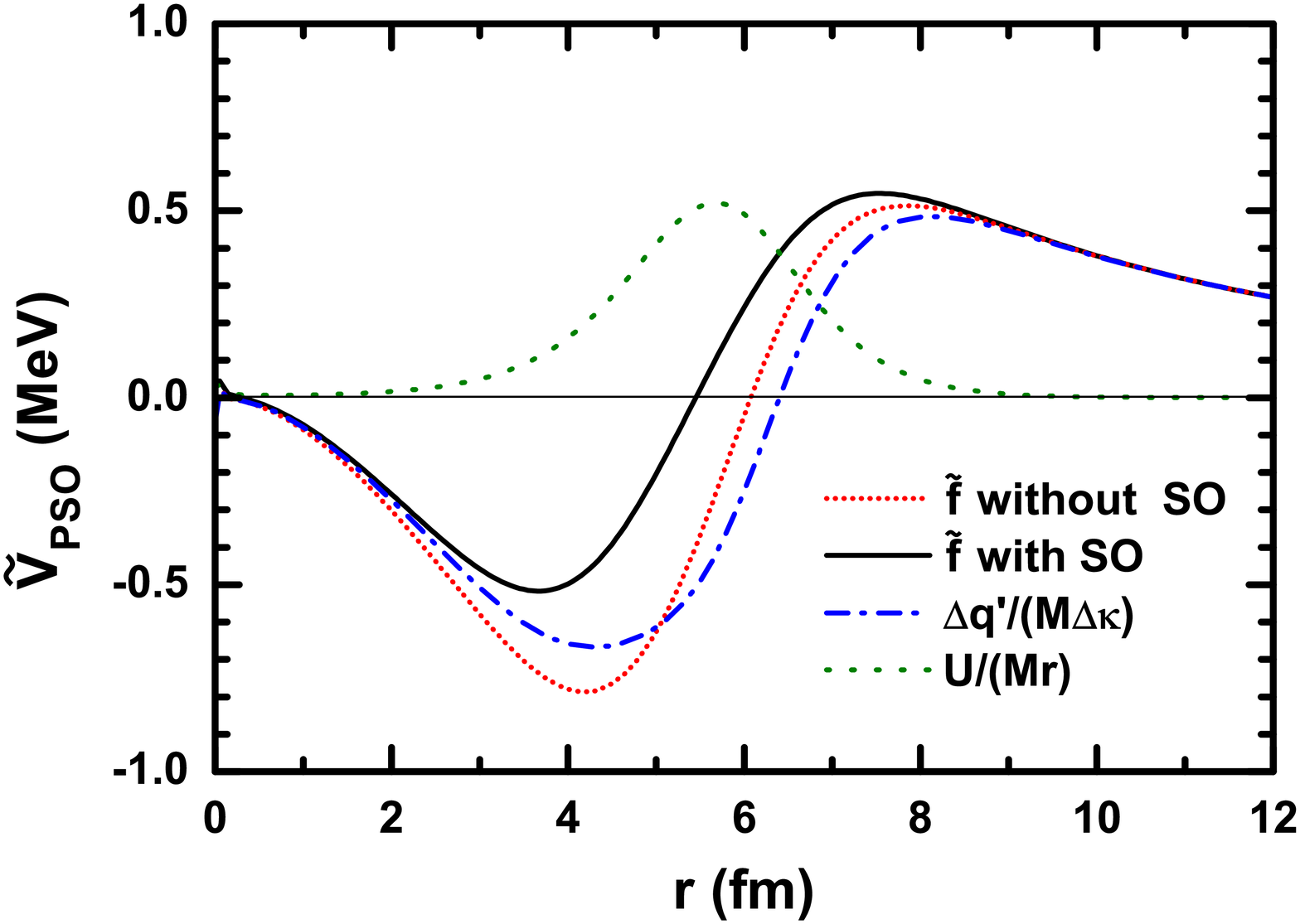}\hspace{2em}
\includegraphics[width=0.45\textwidth]{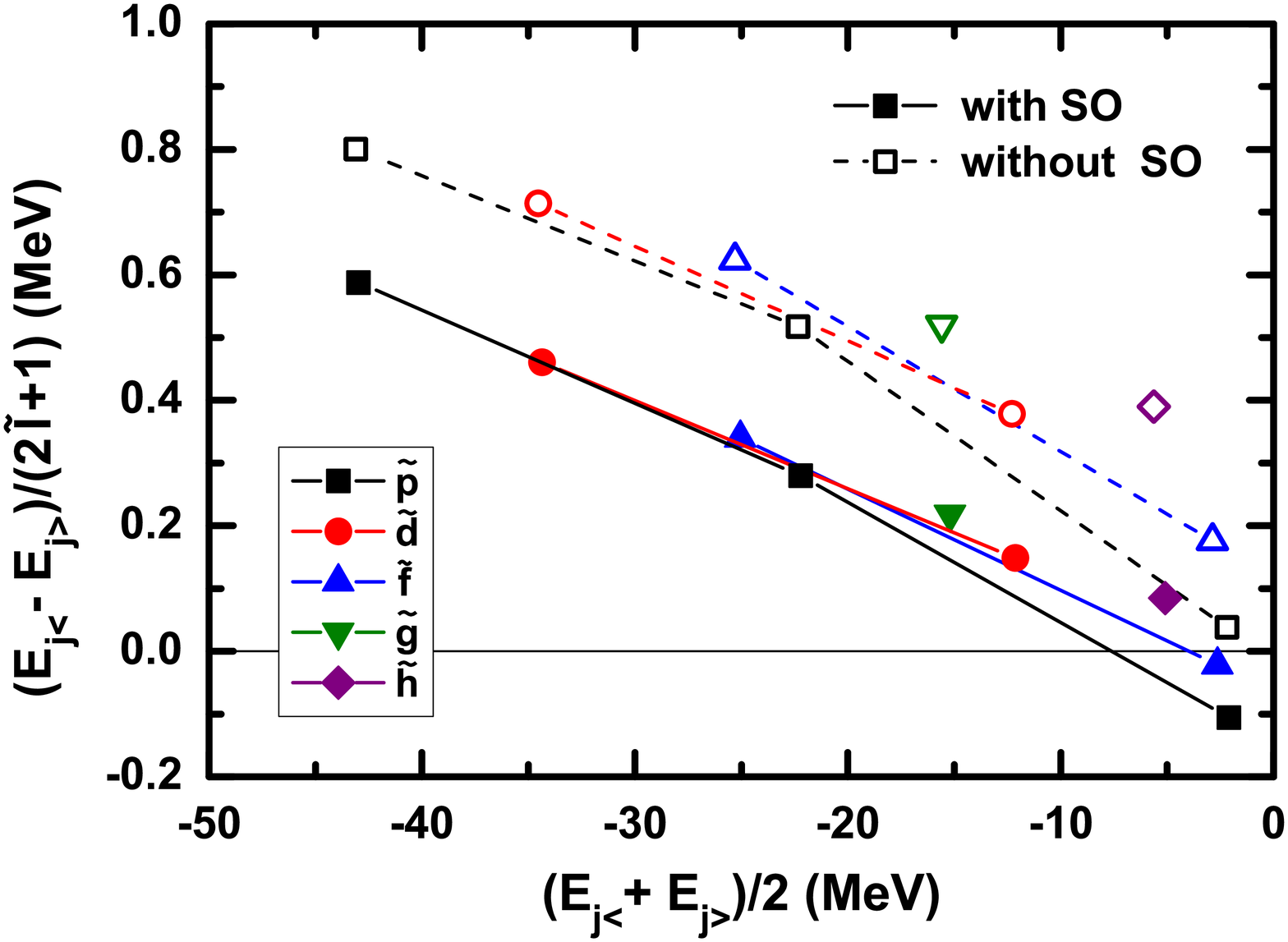}
\caption{(Color online) Left: Pseudospin-orbit potentials $\tilde{V}_{\rm PSO}(r)$ for the $\tilde{f}$ block.
The symmetry-breaking potential obtained with the SO term (solid line) is decomposed into the contributions from the first (dash-dotted line) and second (dotted line) terms on the right-hand side of Eq.~(\ref{Eq:VPSO}).
The symmetry-breaking potential obtained without SO term is shown with short dotted line for comparison.
Right: Reduced pseudospin-orbit splittings $(E_{j_<}-E_{j_>})/(2\tilde{l}+1)$ vs the average single-particle energies $(E_{j_<}+E_{j_>})/2$.
The results obtained with and without SO term are shown with filled and open symbols, respectively.
Taken from Ref.~\cite{Shen2013}.
    \label{Fig1}}
\end{figure}

\section{An example}

In this report, we take the scalar and vector potentials of the Woods-Saxon form from Ref.~\cite{Koepf1991}, and take the neutron potentials in the nucleus $^{132}$Sn as an example.

First of all, the PSS-breaking potentials $\tilde{V}_{\rm PSO}(r)$ for the $\tilde{f}$ block obtained without and with the SO term are shown in the left panel of Fig.~\ref{Fig1}.
These potentials show several special features~\cite{Liang2013,Shen2013}, which are crucial for understanding the PSS in a quantitative way:
(1) they are regular functions of $r$;
(2) their amplitudes directly determine the sizes of reduced PSO splittings $\Delta E_{\rm PSO}\equiv(E_{j_<}-E_{j_>})/(2\tilde l+1)$ according to the perturbation theory;
(3) their shape, being negative at small radius but positive at large radius with a node at the surface region, can explain a general tendency that the PSO splittings become smaller with increasing single-particle energies, and even reverse as approaching the single-particle threshold.

In order to identify the SO effects, the $\tilde{V}_{\rm PSO}(r)$ obtained with the SO term is decomposed into the contributions from the first and second terms on the right-hand side of Eq.~(\ref{Eq:VPSO}), denoted as $\Delta q'/M\Delta\kappa$ and $U/Mr$, respectively.
These two terms can be regarded as the indirect and direct effects of the SO term, respectively, because the former one represents the SO effects on $\tilde{V}_{\rm PSO}(r)$ via the reduced superpotentials $q_\kappa(r)$, while the latter is nothing but the SO potential itself appearing in Eq.~(\ref{Eq:Schrodinger}).
Comparing to the result obtained without SO term, it is found that the indirect effect on $\tilde{V}_{\rm PSO}(r)$ is only around $0.1\sim0.2$~MeV, and eventually results in less influence on $\Delta E_{\rm PSO}$ due to the cancellation between $r<5$~fm and $r>5$~fm regions.
On the other hand, the SO potential $U(r)/Mr$ is always positive with a surface-peak shape.
It substantially raises the $\tilde{V}_{\rm PSO}(r)$, in particular for the surface region.
This systematically reduces the PSO splittings $\Delta E_{\rm PSO}$.

All of these properties are confirmed in the right panel of Fig.~\ref{Fig1}, in which $\Delta E_{\rm PSO}$ for all bound pseudospin doublets are shown as a function of the average single-particle energies $E_{\rm av}=(E_{j_<}+E_{j_>})/2$.
The results obtained with and without SO term are shown with filled and open symbols, respectively.
It is found that the sizes of $\Delta E_{\rm PSO}$ match the amplitudes of $\tilde{V}_{\rm PSO}(r)$.
The decreasing of the PSO splittings with increasing single-particle energies is due to the special shape of $\tilde{V}_{\rm PSO}(r)$.
Last but not least, the SO term reduces the $\Delta E_{\rm PSO}$ systematically by $0.15\sim0.3$~MeV, and this effect can be understood now in a fully quantitative way.

\section{Summary}

Work is now in progress for exploring the origin of PSS and its breaking mechanism by combining the SRG, SUSY quantum mechanics, and perturbation theory.
It is shown that while the spin-symmetry-conserving term appears in the single-particle Hamiltonian $H$, the PSS-conserving term appears naturally in its SUSY partner Hamiltonian $\tilde H$.
The eigenstates of Hamiltonians $H$ and $\tilde H$ are exactly one-to-one identical except for the so-called intruder states.
In such a way, the origin of PSS deeply hidden in $H$ can be traced in its SUSY partner Hamiltonian $\tilde H$.
Furthermore, the perturbative nature of PSS is demonstrated by the perturbation calculations, and the PSS-breaking term can be regarded as a small perturbation on the exact PSS limits, so that the special patterns of the PSO splittings can be interpreted in a quantitative way.

\section*{Acknowledgements}

This work was partly supported by
the RIKEN iTHES Project,
the Grant-in-Aid for JSPS Fellows under Grant No. 24-02201;
the Major State 973 Program No. 2013CB834400;
the NSFC under Grants No. 11105005, No. 11105006, and No. 11175002;
the China Postdoctoral Science Foundation under Grant No. 2012M520101;
and the Research Fund for the Doctoral Program of Higher Education under Grant No. 20110001110087.

\section*{References}


\providecommand{\newblock}{}

\end{document}